\documentstyle[12 pt]{article}

\begin{document}

\bigskip
\begin{center}
{ \Large {\bf Quantum mechanics without spacetime IV }}\\
{\large {\bf - A noncommutative Hamilton-Jacobi equation - }}

\bigskip

{{\large
{\bf T. P. Singh\footnote{e-mail address: tpsingh@tifr.res.in} }}}

\smallskip

{\it Tata Institute of Fundamental Research,}\\
{\it Homi Bhabha Road, Mumbai 400 005, India.}
\vskip 0.5cm
\end{center}

\bigskip

\begin{abstract}

\noindent It has earlier been argued that there should exist a formulation
of quantum mechanics which does not refer to a background spacetime. In this
paper we propose that, for a relativistic particle, such a formulation is
provided by a noncommutative generalisation of the Hamilton-Jacobi equation.
If a certain form for the metric in the noncommuting coordinate system is
assumed, along with a correspondence rule for the commutation relations,
it can be argued that this noncommutative Hamilton-Jacobi equation is 
equivalent to standard quantum mechanics. 

\end{abstract}

\section{Introduction}

The standard formulation of quantum mechanics assumes the 
existence of a
classical background spacetime. However, such a classical spacetime becomes
possible only because the present universe is dominated by classical matter
fields. In the complete absence of such classical matter fields it becomes essential
to formulate the rules of quantum mechanics without reference to a background
classical spacetime \cite{qm1}, \cite{carlip}.

A possible reformulation is in the language of noncommutative geometry, as has
been motivated in two preliminary papers \cite{qm1, qm2}. The essential idea
is as follows. We assume that, in addition to the standard Minkowski and 
curvilinear coordinate systems, there exist in nature, noncommuting coordinate
systems. That is, in such a coordinate system the coordinates do not commute
with each other. We also assume that a description of quantum
mechanics which is independent of a classical spacetime can be given using 
such noncommuting coordinates.

Consider first the case of a single relativistic particle, having a mass $m$ much less than the Planck mass $m_{Pl}$. As has been argued 
earlier \cite{qm1}, in this case one can legitimately neglect the 
gravitational field of the particle. 
The nature of the dynamics for $m\ll m_{Pl}$, in the absence of a background classical spacetime, is the subject
of this paper. It is assumed that there are many different possible 
noncommuting coordinate systems, which could all be equivalently used to
describe dynamics (in the same spirit as Lorentz invariance, or general
covariance). This equivalence is referred to as automorphism invariance - a
natural generalisation when one makes a transition from Riemannean geometry
to a noncommutative geometry \cite{connes}.

When we examine the dynamics of this quantum mechanical particle from our
vantage point, we are of course using standard commuting spacetime coordinates to 
describe its dynamics. It thus has to be shown that the fundamental
description of the dynamics in terms of noncommuting coordinates reduces to
quantum mechanics as we know it, when that dynamics is examined from a 
commuting coordinate system. The introduction of such commuting coordinate systems becomes possible because of the dominant presence of classical matter fields, in
much the same way that Minkowski coordinate systems can be used very accurately in asymptotic regions, even though we well know that the real Universe is curved. In this spirit one is proposing that in the absence of a classical background, an appropriate description of
quantum mechanics is via the use of noncommuting coordinate systems.

Here, without any pretense at achieving the rigor of noncommutative geometry,
we construct a model for the noncommutative dynamics of a 
relativistic particle, and suggest how the model could be related to standard
quantum mechanics.

\section{A noncommutative Hamilton-Jacobi equation}

Let us start by considering the relativistic Schrodinger equation 
\cite{schiff} for a
particle in 2-d spacetime
\begin{equation}-\hbar^{2}
\left({\partial^{2}\over\partial t^{2}}-{\partial^{2}\over\partial x^{2}}\right)\psi=m^{2}\psi\label{kg}
\end{equation}
which, after the substitution $\psi=e^{iS/\hbar}$, becomes
\begin{equation}
\left({\partial{S}\over \partial t}\right)^{2}-\left({\partial{S}\over \partial x}\right)^{2}
-i\hbar\left({\partial^{2}S\over\partial t^{2}}-{\partial^{2}S\over\partial x^{2}}\right)=m^{2}\label{hjc}
\end{equation}
This equation can further be written as
\begin{equation}
p^{\mu}p_{\mu} + i\hbar {\partial p^{\mu}\over \partial x^{\mu}} = m^{2}
\label{hjp}
\end{equation}
where we have defined
\begin{equation}
p^{t}=-{\partial S\over \partial t}, \qquad p^{x} = 
{\partial S \over \partial x}
\label{pmoo}
\end{equation}
and the index $\mu$ takes the values $1$ and $2$.

Equation (\ref{hjc}) could be thought of as a generalisation of the
classical Hamilton-Jacobi equation to the quantum mechanical case [also
for reasons which will become apparent as we proceed], where the `action' 
function $S(x,t)$ is now complex. Evidently, in (\ref{hjp}) the $\hbar$
dependent terms appear as corrections to the classical term 
$p^{\mu}p_{\mu}$. We chose to consider the relativistic case, as
opposed to the non-relativistic one, only because the available space-time
symmetry makes the analysis more transparent.

Taking clue from the form of Eqn. (\ref{hjp}) we now suggest a model for the
dynamics, in the language of two noncommuting coordinates 
$\hat{x}$ and $\hat{t}$. We ascribe to the particle a `momentum' $\hat{p}$,
having two components $\hat{p}^{t}$ and $\hat{p}^{x}$, which do not
commute with each other. The noncommutativity of these momentum components
is assumed to be a consequence of the noncommutativity of the coordinates,
as the momenta are defined to be the partial derivatives of the complex action
$S(\hat{x},\hat{t})$, with respect to the corresponding noncommuting coordinates. 

We further assume that
the coordinates $\hat{x}$ and $\hat{t}$ describe the non-commutative version
of Minkowski space and that the noncommutative flat metric is
\begin{eqnarray}
\label{ncfm}  
\hat{\eta}_{\mu\nu} = \left(\begin{array}{cc}
                      1 & 1 \\
                      -1 & -1 \end{array} \right)
\end{eqnarray}
The determinant is zero - it is not obvious that this is really an
unsatisfactory feature in the present context (we remark further on this, 
below). This metric is obtained by adding an antisymmetric component to the 
Minkowski metric.

We now propose that the background independent description of the quantum
dynamics is given by the equation
\begin{equation}
\hat{p}^{\mu}\hat{p}_{\mu} = m^{2}
\label{nchj}
\end{equation}
Here, $\hat{p}_{\mu}=\hat{\eta}_{\mu\nu}\hat{p}^{\mu}$ is well-defined. Written
explicitly, this equation becomes
\begin{equation}
(\hat{p}^{t})^{2}-(\hat{p}^{x})^{2} + 
\hat{p}^{t}\hat{p}^{x} - \hat{p}^{x}\hat{p}^{t}  = m^{2}
\label{nce}
\end{equation} 
Eqn. (\ref{nchj}) appears an interesting and plausible proposal for the 
dynamics, because it generalizes the corresponding special relativistic 
equation to the noncommutative case. The noncommutative Hamilton-Jacobi
equation is constructed from (\ref{nce}) by defining the momentum as gradient
of the complex action function.
 
Of course we now need to ask if this 
dynamics looks the same as quantum mechanics, when seen from our classical 
spacetime. It should be apparent that a description of the 
noncommutative dynamics from the vantage point of a classical spacetime, is 
actually an approximate one. It is like trying to describe curved space 
dynamics using a flat spacetime metric - errors would be introduced, which 
have to be rectified by adding correction terms. 

Thus, we propose the following  rule for the 
transformation of the expression on the left of (\ref{nce}), when it is
seen from our standard commuting coordinate system:
\begin{equation}
(\hat{p}^{t})^{2}-(\hat{p}^{x})^{2} + 
\hat{p}^{t}\hat{p}^{x} - \hat{p}^{x}\hat{p}^{t}  = ({p}^{t})^{2}-({p}^{x})^{2}
 + i\hbar {\partial p^{\mu}\over \partial x^{\mu}}
\label{nceq}
\end{equation}
Here, $p$ is the `momentum' of the particle as seen from the commuting
coordinate system, and is related to the complex action by Eqn. (\ref{pmoo}).
This equality should be understood as an equality between the two 
equivalent equations
of motion for the complex action function $S$ - one written in the
noncommuting coordinate system, and the other written in the standard
commuting coordinate system.

The idea here is that by using the Minkowski metric of ordinary spacetime 
one does not correctly measure the length of the `momentum' vector, because 
the noncommuting off-diagonal part is missed out. The last, 
$\hbar$ dependent term in (\ref{nceq}) provides the correction - the origin
of this term's relation to the commutator $\hat{p}^{t}\hat{p}^{x} - \hat{p}^{x}\hat{p}^{t}$ remains to be understood. If the relation (\ref{nceq}) holds, 
there is equivalence between the background independent description 
(\ref{nchj}) and standard quantum dynamics given by (\ref{hjp}).

The proposal proceeds along analogous lines for four-dimensional spacetime.
The metric $\hat{\eta}_{\mu\nu}$ is defined by adding an antisymmetric part 
(all entries of which are $1$ and $-1$) to the Minkowski metric, and the 
off-diagonal contribution on the left-hand side of (\ref{nceq}) is to be set 
equal to $i\hbar {\partial p^{\mu}/ \partial x^{\mu}}$ on the right hand
side.

\section{Discussion}

One could consider arriving at the noncommutative dynamics by a 
different route. Suppose we demand that the allowed class of noncommutative
coordinate transformations are those which leave the line-element
\begin{equation}
ds^{2}= \hat{\eta}_{\mu\nu} 
        d\hat{x}^{\mu}d\hat{x}^{\nu} = d\hat{t}^{2}-d\hat{x}^{2}+ 
d\hat{t} d\hat{x} - d\hat{x} d\hat{t} 
\label{le}
\end{equation}
invariant. The noncommutative flat metric is given by (\ref{ncfm}).
This will be `flat noncommutative spacetime automorphism 
invariance' - a generalisation of Lorentz invariance.

This line-element is left invariant by the Lorentz transformation
\begin{equation}
\hat{x}'=\gamma({\hat{x}-\beta\hat{t}}),\qquad 
\hat{t}'=\gamma({\hat{t}-\beta\hat{x}})
\end{equation}
where $\gamma=(1-\beta^{2})^{-1/2}$. In the commutative limit, $\beta$ has the interpretation of velocity: $\beta=v/c$. In the noncommutative case, $\beta$ should be thought of as defining a rotation in the non-commutative space by an angle $\theta$ defined by $\beta=\tanh\theta$.

It appears reasonable to expect that, since this class of transformations
generalises Lorentz transformations to the noncommutative case, the dynamics
should now be given by (\ref{nchj}), which is a generalisation of the
corresponding equation in special relativity. Hence the question posed here
is: is quantum mechanics the same as the mechanics obtained by generalising
special relativity to a noncommutative spacetime? The question as to
exactly what is the form of the commutation relation for the noncommuting coordinates is still open.

A possible structure for the commutation relations might be as follows. We
assume fundamental relations of the form
\begin{equation}
[\hat{t},\hat{x}]=iL_{Pl}^{2}, \qquad [\hat{p}^{t},\hat{p}^{x}]=iP_{Pl}^{2}
\label{fcr}
\end{equation}
where $P_{Pl}$ is Planck-momentum $(=m_{Pl}c)$. This would suggest uncertainty
relations of the kind
\begin{equation}
\Delta\hat{t}\ \Delta\hat{x}\sim L_{Pl}^{2}, \qquad 
\Delta\hat{p}^{t}\ \Delta\hat{p}^{x}\sim P_{Pl}^{2}
\end{equation}
and hence
\begin{equation}
\Delta\hat{t}\ \Delta\hat{x}\ \Delta\hat{p}^{t}\ \Delta\hat{p}^{x}
\sim L_{Pl}^{2}P_{Pl}^{2}=\hbar^{2}
\end{equation}
which by symmetry suggests that
\begin{equation}
\Delta\hat{t}\ \Delta\hat{p}^{t}
\sim \hbar,\qquad
\Delta\hat{x}\ \Delta\hat{p}^{x}
\sim \hbar
\end{equation}
If this last uncertainty relation is invariant under transformation
from the noncommuting coordinate system to the commuting one, we might be
able to understand the standard quantum commutation relation $[q,p]=i\hbar$ 
as having its origin in the fundamental commutation relations (\ref{fcr}). 
While its generally regarded as implausible that quantum mechanics could be a
consequence of ``Planck-scale'' physics, we feel it may not be completely
unreasonable to consider such a connection.  

One could also try to derive the non-commutative Hamilton-Jacobi equation as a consequence of the variational principle
\begin{equation}
\delta S = mc\ \delta \int ds = 0
\end{equation}
where $ds$ is defined by the non-commutative flat line-element (\ref{le}). Work
is in progress to address these unresolved issues, as well as the 
multi-particle case, the non-relativistic limit, and a detailed examination of
the properties of a noncommutative special relativity. 

In the 4-d case, the noncommutative line-element analogous to (\ref{le}) is
not invariant under a Lorentz transformation, thus suggesting that here the
appropriate coordinate transformation is a generalization of Lorentz
transformations. 

When one assumes $m\ll m_{Pl}$, as was done here, one is neglecting gravity.
Allowing for the mass to be comparable to Planck mass implies the introduction
of curvature and a generalisation of the noncommutative flat metric to a
noncommutative curved metric. In particular, the off-diagonal antisymmetric components of the metric will become mass-dependent, and the determinant will no
longer be zero. Thus the vanishing of the determinant is strictly only a
theoretical situation, and not an actual one. In reality, for $m\ll m_{Pl}$
the determinant is very close to zero, but not exactly zero.

The introduction of curvature and departure from the noncommutative flat 
metric brings into the picture general coordinate transformations of
noncommuting coordinates (thus generalising general covariance to automorphism
invariance). In such a case, one expects significant departures from the 
Schrodinger equation when $m\approx m_{Pl}$, an example of which was discussed
in \cite{qm3}. The experimental search for such a departure would go a long way
in establishing contact between experiment and theoretical quantum gravity. 

\bigskip

\noindent{\bf Acknowledgments}: I would like to thank Sashideep Gutti, N. Kumar, Joseph Samuel, R. Srikanth, Sumati Surya and Cenalo Vaz for useful discussions. It is a pleasure to also acknowledge the warm hospitality of the Raman Research Institute, Bangalore.

\end{document}